\documentclass[twocolumn,showpacs,preprintnumbers,amsmath,amssymb,prl,superscriptaddress]{revtex4-1}
\usepackage{bbm}
\usepackage{mathrsfs}
\usepackage{graphicx}
\usepackage{dcolumn}
\usepackage{bm}
\usepackage{amsmath}
\usepackage{amsfonts}
\usepackage{color}
\usepackage[colorlinks=true,citecolor=blue,anchorcolor=blue]{hyperref}
\usepackage{floatrow}

\begin{document}

\title{Large dynamical axion field in topological antiferromagnetic insulator Mn$_2$Bi$_2$Te$_5$}
\author{Jinlong Zhang}
\affiliation{State Key Laboratory of Surface Physics, Department of Physics, Fudan University, Shanghai 200433, China}
\author{Dinghui Wang}
\affiliation{National Laboratory of Solid State Microstructures, School of Physics, Nanjing University, Nanjing 210093, China}
\author{Minji Shi}
\affiliation{National Laboratory of Solid State Microstructures, School of Physics, Nanjing University, Nanjing 210093, China}
\author{Tongshuai Zhu}
\affiliation{National Laboratory of Solid State Microstructures, School of Physics, Nanjing University, Nanjing 210093, China}
\author{Haijun Zhang}
\thanks{zhanghj@nju.edu.cn}
\affiliation{National Laboratory of Solid State Microstructures, School of Physics, Nanjing University, Nanjing 210093, China}
\affiliation{Collaborative Innovation Center of Advanced Microstructures, Nanjing 210093, China}
\author{Jing Wang} 
\thanks{wjingphys@fudan.edu.cn}
\affiliation{State Key Laboratory of Surface Physics, Department of Physics, Fudan University, Shanghai 200433, China}
\affiliation{Collaborative Innovation Center of Advanced Microstructures, Nanjing 210093, China}
\affiliation{Institute for Nanoelectronic Devices and Quantum Computing, Fudan University, Shanghai 200433, China}

\begin{abstract}
The dynamical axion field is a new state of quantum matter where the magnetoelectric response couples strongly to its low-energy magnetic fluctuations. It is fundamentally different from an axion insulator with a static quantized magnetoelectric response. The dynamical axion field exhibits many exotic phenomena such as axionic polariton and axion instability. However, these effects have not been experimentally confirmed due to the lack of proper topological magnetic materials. Here by combining analytic models and first-principles calculations, we predict a series of van der Waal layered Mn$_2$Bi$_2$Te$_5$-related topological antiferromagnetic materials could host the long-sought dynamical axion field with a topological origin. We also show a large dynamical axion field can be achieved in antiferromagnetic insulating states close to the topological phase transition. We further propose the optical and transport experiments to detect such a dynamical axion field. Our results could directly aid and facilitate the search for topological-origin large dynamical axion field in realistic materials.
\end{abstract}

\date{\today}


\maketitle

\emph{Introduction.}
Topological phenomena in physical systems are determined by some topological structure and are thus usually universal and robust against perturbations~\cite{thouless1998}. The interplay between band topology and magnetic order brings the opportunity to realize a large family of exotic topological phenomena~\cite{tokura2019,qi2008,hasan2010,qi2011,qi2009b,wang2017c,li2010,yu2010,maciejko2010,tse2010,nomura2011,wang2015b,morimoto2015,okada2016,wul2016,dziom2017,mogi2017,mogi2017a,xiao2018,zhang2019,bermudez2010,lee2015,wieder2018,gui2019,xu2019,wan2012,turner2012,wangzhong2013,you2016,gooth2019}. The electromagnetic response of a three-dimensional insulator is described by the topological $\theta$ term $\mathcal{S}_{\theta}=(\theta/2\pi)(e^2/h)\int d^3xdt\mathbf{E}\cdot\mathbf{B}$~\cite{qi2008}, together with the ordinary Maxwell action. Here $\mathbf{E}$ and $\mathbf{B}$ are the conventional electromagnetic fields inside the insulator, $e$ is the charge of an electron, $h$ is Planck's constant, and $\theta$ is the dimensionless pseudoscalar parameter describing the insulator, which refers to axion field in particle physics and could explain the missing dark matter~\cite{peccei1977,wilczek1987}. Under periodic boundary conditions, all physical quantities depends on $\theta$ only module $2\pi$. While $\mathcal{S}_{\theta}$ generically breaks time-reversal symmetry $\mathcal{T}$ and parity $\mathcal{P}$, both symmetries are conserved at $\theta=0$ and $\theta=\pi$. Axion insulators and topological insulators (TIs) have $\theta=\pi$ which is protected by $\mathcal{P}$ and $\mathcal{T}$, respectively. TI can be connected continuously to trivial insulator defined by $\theta=0$, only by $\mathcal{T}$-breaking perturbations. $\mathcal{S}_{\theta}$ with the static $\theta=\pi$ leads to image monopole~\cite{qi2009b} and quantized topological magnetoelectric effect (TME)~\cite{qi2008,nomura2011,wang2015b,morimoto2015,karch2009,mulligan2013,rosenow2017,franz2010,liuzc2020}.

In a uniform TI material, the parameter $\theta$ is time and position independent, and $\mathcal{S}_{\theta}$ is only a surface term. When the antiferromagnetic (AFM) long-range order is developed, $\theta$ becomes a bulk dynamical field from magnetic fluctuations and taking continuous values from $0$ to $2\pi$, and $\mathcal{S}_{\theta}$ becomes a bulk term. Such magnetic materials with dynamical axion fields (DAFs) leads to many exotic physical effects such as axionic polariton~\cite{li2010}, axion instability induced nonlinear electromagnetic effect~\cite{ooguri2012,imaeda2019} and so on~\cite{wang2011a,wang2016a,sekine2016,sekine2016b,taguchi2018}. The realization of a DAF require a proper coupling between electrons and magnetic fluctuations. The conventional magnetoelectric material Cr$_2$O$_3$ has diagonal but anisotropic magnetoelectric response, while the pseudosclar part may exhibit DAF~\cite{hehl2008,wangjl2019}. However, the axion field, if exists in Cr$_2$O$_3$, is expected to be quite weak as discussed below. In this work, we combine analytic models and first-principles calculations, to predict Mn$_2$Bi$_2$Te$_5$ family materials as potential candidates hosting a large DAF. One direct consequence of DAF is chiral magnetic effect (CME), where an alternating electric current is generated by static magnetic fields from DAF induced by AFM resonance, which is absent in an axion insulator.

\emph{Axion electrodynamics.} As $\theta$ is odd under $\mathcal{T}$ and $\mathcal{P}$ operation, only $\mathcal{T}$- and $\mathcal{P}$-breaking perturbations can induce a change of $\theta$. So the fluctuations of AFM order with $\mathcal{T}, \mathcal{P}$-breaking can induce $\delta\theta(\mathbf{r},t)$~\cite{li2010}. A large $\delta\theta(\mathbf{r},t)$ is expected when the magnetic fluctuation is strong. However, this is not sufficient. Here we start with a simple Dirac model describing an insulator and show the relation between $\theta(\mathbf{r},t)$ and AFM order, to see how a large DAF can be achieved, which is generic for any system supporting axionic excitation. The generic effective Hamiltonian is~\cite{li2010}
\begin{equation}\label{Dirac}
\mathcal{H}_{\mathrm{Dirac}}=\epsilon_0(\mathbf{k})+\sum\limits_{a=1}^5d_a(\mathbf{k})\Gamma^a,
\end{equation}
where $d_{1,2,3,4,5}(\mathbf{k})=(Ak_x,Ak_y,Ak_z,m_4(\mathbf{k}),m_5)$, and $m_4(\mathbf{k})=m+Bk^2$. For simplicity, we neglect the particle-hole asymmetry $\epsilon_0(\mathbf{k})$ and set the velocity $A$ and curvature $B$ along three axes to be isotropic, which does not affect the physics we discuss here. $\Gamma^a$ are Dirac matrices satisfying the Clifford algebra $\{\Gamma^a,\Gamma^b\}=2\delta_{ab}$ and $\Gamma^5=\Gamma^1\Gamma^2\Gamma^3\Gamma^4$. Also $\mathcal{T}\Gamma^a\mathcal{T}^{-1}=-\Gamma^a$ and $\mathcal{P}\Gamma^a\mathcal{P}^{-1}=-\Gamma^a$ for $a=1, 2, 3, 5$, and $\mathcal{T}\Gamma^4\mathcal{T}^{-1}=\mathcal{P}\Gamma^4\mathcal{P}^{-1}=\Gamma^4$. $m_4$ is the $\mathcal{T}$-invariant mass, and $m_5$ is the $\mathcal{T}, \mathcal{P}$-breaking mass proportional to AFM order. When $m_5=0$, the model describes a TI (or trivial insulator) if $m/B<0$ (or $m/B>0$). $m_5$ leads to a correction to $\theta$ to the linear order, the value of $\theta$ can be calculated from the momentum-space Chern-Simons form~\cite{qi2008,li2010,essin2009}. Namely, $\theta(\mathbf{r},t)=\theta_0+\delta\theta(\mathbf{r},t)$, where $\theta_0$ and $\delta\theta(\mathbf{r},t)$ is the static and dynamical part of the axion field, respectively, and
\begin{equation}
\delta\theta(\mathbf{r},t)=\delta m_5(\mathbf{r},t)/g,
\end{equation}
where $\delta m_5(\mathbf{r},t)$ is proportional to the AFM fluctuations, and $1/g=\partial\theta/\partial m_5$. Therefore, not only AFM fluctuations but also the nonmagnetic mass $m_4$ (through $1/g$) could affect $\delta\theta$. There are other leading order $\mathcal{T}, \mathcal{P}$-breaking terms $\sum_{i=1}^3m_i\Gamma^i$ which only give rise to higher-order contributions to $\theta$, and thus are neglected here~\cite{supple}.

\begin{figure}[t]
\begin{center}
\includegraphics[width=3.4in,clip=true]{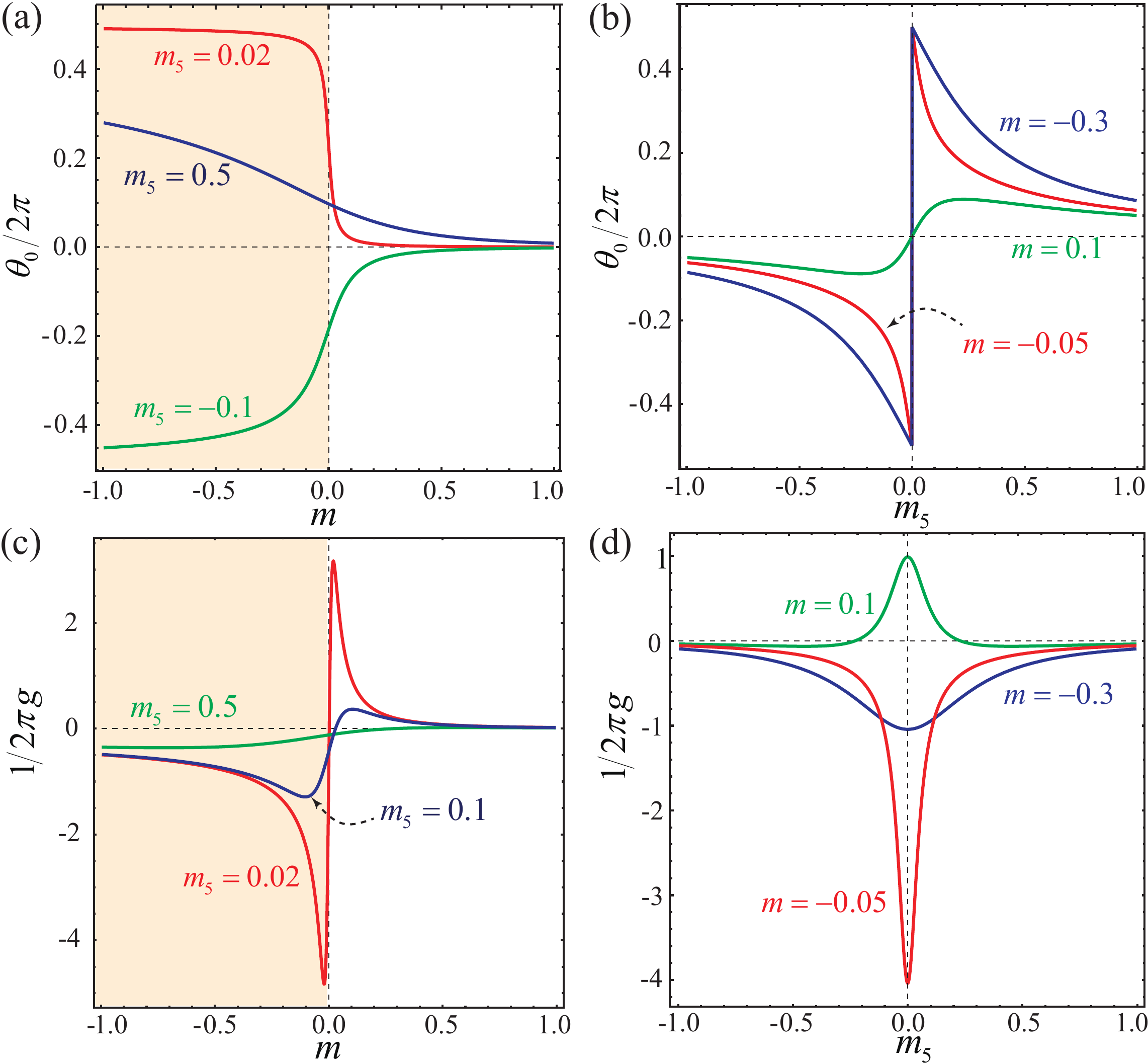}
\end{center}
\caption{$\theta_0$ and $1/g$ for the massive Dirac model. (a) $\theta_0$ vs $m$. (b) $\theta_0$ vs $m_5$. $\theta_0=\pm\pi$ when $m_5\rightarrow0^{\pm}$ and $m<0$, and $\theta_0=0$ when $m_5=0$ and $m>0$. (c) $1/g$ vs $m$. (d) $1/g$ vs $m_5$. The other parameters are $A=1.0$, $B=0.5$, and the arbitrary units are chosen for $1/g$, $m$ and $m_5$.}
\label{fig1}
\end{figure}

Typical values of $\theta_0$ as a function of $m$ and $m_5$ are calculated in Fig.~\ref{fig1}(a) and~\ref{fig1}(b), respectively. As expected, $\theta_0$ deviates from $\pm\pi$ ($m/B<0$) or $0$ ($m/B>0$) when $m_5$ is nonzero. $\theta$ is an odd function of $m_5$, this implies $g$ is an even function of $m_5$ as shown in Fig.~\ref{fig1}(d), $1/|g|$ is largest when $m_5=0$. From Fig.~\ref{fig1}(c), for $m_5$ very close to zero, $1/g$ have a dip (peak) when $m\rightarrow0^-$ ($m\rightarrow0^+$), and $1/|g|$ at the dip is larger than at the peak. Both the dip and peak in $1/g$ are further suppressed when $m_5$ increases, and finally are smeared out when $m_5$ is large. In the latter case, $1/|g|$ increases when $m$ becomes inverted.  Here we point out, the asymmetry dependence of $1/|g|$ on positive and negative $m$ is from the nonzero $B$. In the limit of $m_5=B=0$, $1/g\propto1/m$. Thus $\delta\theta$ is very sensitive to a small change in $m_5$ when $m$ is close to zero. This suggests that, to get a large $\delta\theta(\mathbf{r},t)$, both large $\delta m_5$ and $1/|g|$ are required, where the latter is achieved when $m_5\rightarrow0$ and $m$ approaches zero (preferably from the topological \emph{nontrivial} side). This further indicates a most likely promising strategy  for a large DAF: as long as $\theta$ is well defined, a $\mathcal{T,P}$-broken insulator which is close to topological phase transition with a small inverted $m$ and is also close to paramagnetic to AFM transition with $m_5\approx0$, where the magnetic fluctuation is expected to be strong. In general, in a specific AFM material, the magnetic fluctuation and AFM order are determined by the intrinsic screened Coulomb interaction between electrons, and thus the amplitude of $\delta m_5$ and $m_5$ are fixed. Therefore, to search for large DAF in practice, intrinsic AFM TIs (having a weak AFM order and is) near the topological phase transition are preferred (tunable $m_4$ by spin-orbit coupling). This also explain that the DAF is too small to be observed in a conventional AFM insulator e.g. Cr$_2$O$_3$, which is far away from topological phase transition.

\begin{figure}[t]
\begin{center}
\includegraphics[width=3.4in,clip=true]{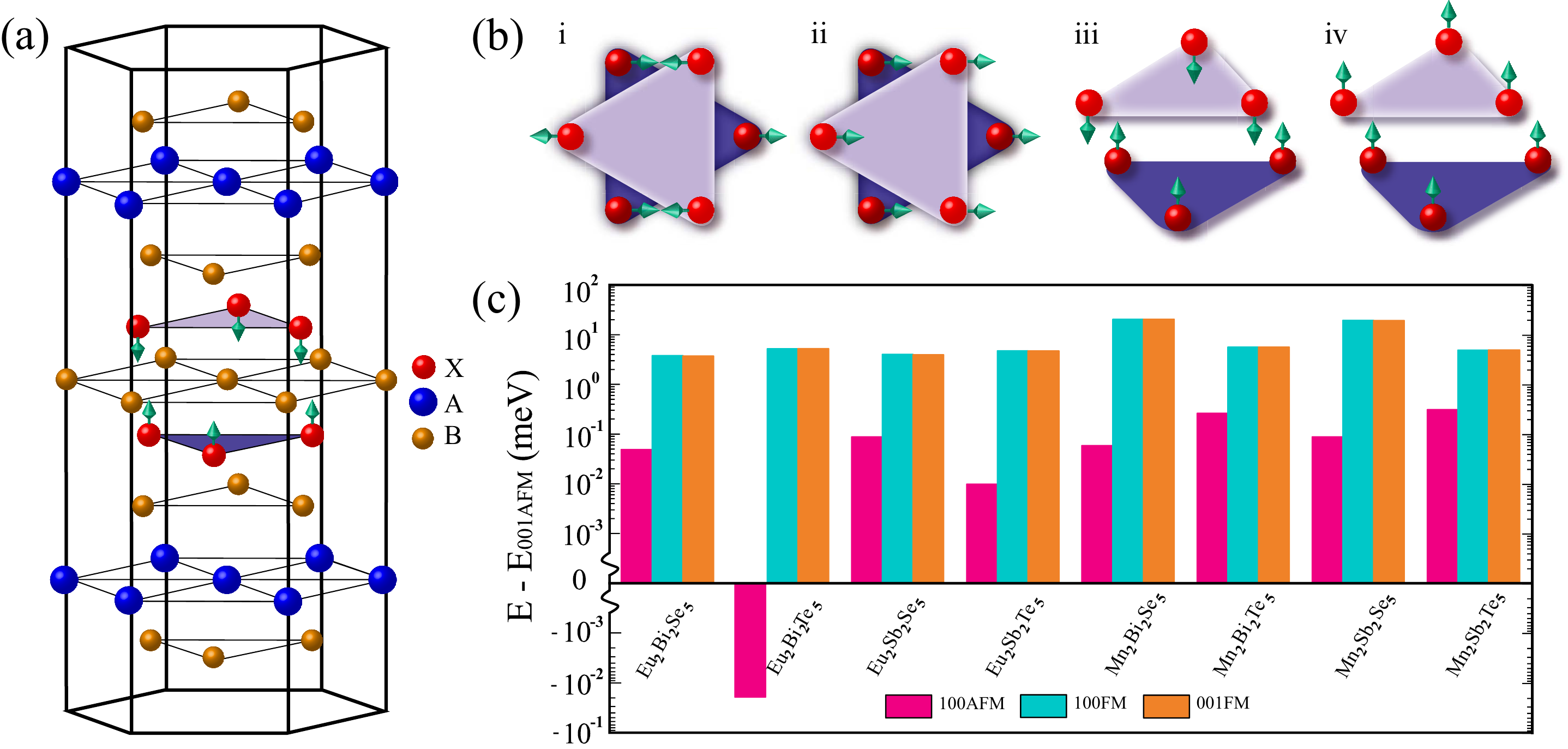}
\end{center}
\caption{Crystal structure and magnetic configurations. (a) The crystal structure of X$_2$A$_2$B$_5$ materials. Each layer has a triangle lattice. The green arrows denote the spin moments of X atom. (b) Sketch of different magnetic configurations. i/iii present 100AFM/001AFM spin moments, and ii/iv present 100FM/001FM spin moments, respectively. (c) The magnetic anisotropy, where the total energy of 001AFM state is set to be zero as reference.}
\label{fig2}
\end{figure}

\emph{Material candidates.} The recent discovery of intrinsic magnetic TIs in MnBi$_2$Te$_4$ family materials~\cite{zhang2019,li2019,otrokov2019,gong2019,deng2020,liu2020} rekindled our hope for searching DAF in intrinsic AFM TIs. The symmetry guiding principle is to find $\mathcal{T},\mathcal{P}$-breaking but $\mathcal{PT}$-conserving TIs. The AFM order in MnBi$_2$Te$_4$ has no contribution to $\delta\theta(\mathbf{r},t)$ due to conserved $\mathcal{P}$. However, a class of ternary chalcogenides materials X$_2$A$_2$B$_5$, also written as (XB)$_2$(A$_2$B$_3$)$_1$, with X=Mn/Eu, A=Sb/Bi, B=Se/Te, satisfy the symmetry requirement for DAF. In the following, we take Mn$_2$Bi$_2$Te$_5$ as an example, which has been successfully synthesized in experiments recently~\cite{lv2020}. It has a layered rhombohedral crystal structure, shown in Fig.~\ref{fig2}(a), with the space group $P\overline{3}m1$ (No. 164) if the spin moments of Mn are ignored. It consists of nine-atom layers (e.g., Te1-Bi1-Te2-Mn1-Te3-Mn2-Te4-Bi2-Te5) arranged along the trigonal $z$-axis with the ABC-type stacking, known as a nonuple layer (NL). The coupling between different NLs is the van der Waals type. Once the AFM order is considered, $\mathcal{T},\mathcal{P}$ are broken.

First-principles calculations are employed to investigate the electronic structures, and the detailed methods can be found in Supplemental Material~\cite{supple}. We find that each Mn atom tends to have half-filled $d$ orbitals with a magnetic moment $\mathbf{S}=5\mu_B$, and the orbital moment is quenched $\mathbf{L}=0$. The total energy calculations for different collinear magnetic structures are given in Fig.~\ref{fig2}(c). Here the non-collinear magnetic structure is not considered due to vanishing Dzyaloshinski-Moriya interaction. In fact, the non-collinear magnetic configuration was previously found to have higher energy~\cite{hou2019}. The $A$-type AFM phase with the out-of-plane easy axis, denoted as 001AFM [seen in Fig.~\ref{fig2}(b)], is the magnetic ground state for most X$_2$A$_2$B$_5$ family materials, except Eu$_2$Bi$_2$Te$_5$. The in-plane $A$-type AFM phase (100AFM) has a slightly higher total energy than that of 001AFM. The FM order is not favored. The magnetic anisotropy of Eu$_2$Bi$_2$Te$_5$ is quite weak because of the local 4$f$ orbital, and a small external magnetic field can easily tune 100AFM into 001AFM. The Goodenough-Kanamori rule is the key to have FM order in each Mn layer. Two nearest Mn atoms are connected through Te atom with the bond ``Mn-Te-Mn'', whose bonding angle is $\sim$ 95 degree, so the superexchange interaction is expected to induce the in-plane FM order.

\begin{figure}[htbp]
\begin{center}
\includegraphics[width=3.4in,clip=true]{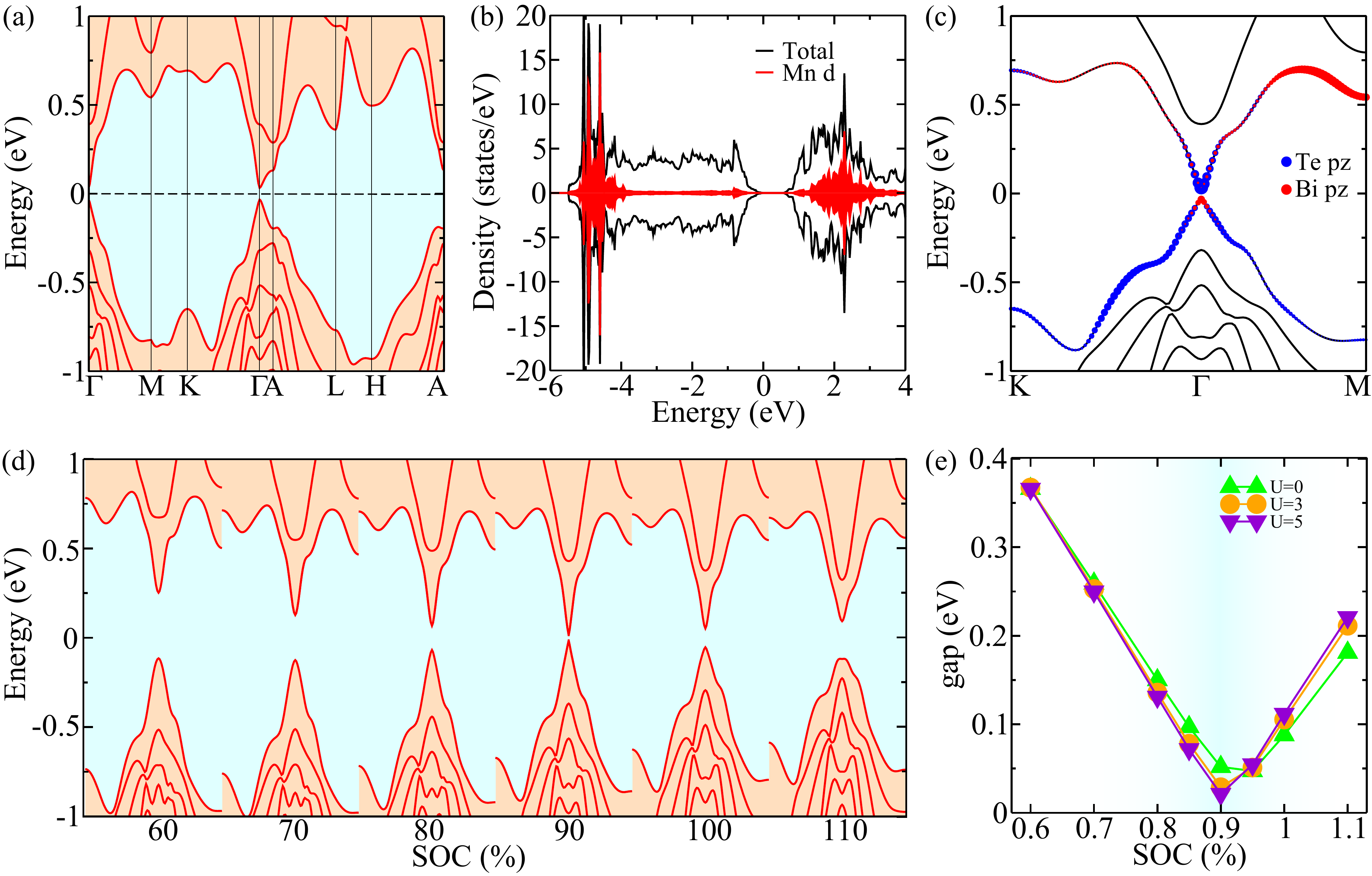}
\end{center}
\caption{Electronic structure. (a) The bulk band structure of 001AFM Mn$_2$Bi$_2$Te$_5$ with SOC. The Fermi level (dashed line) is set as zero. (b) DOS without SOC. The black and red lines present total DOS and the projected DOS for Mn $d$-orbitals, respectively. (c) The fat band structures. The blue and red dots denote the character of Te and Bi $p_z$-orbitals, respectively. (d) The evolution of the band structure with SOC. (e) The evolution of the energy gap with SOC for different onsite $U$ in the LDA+U calculations.}
\label{fig3}
\end{figure}

The 001AFM state breaks $\mathcal{T}$ and $\mathcal{P}$ but preserves $\mathcal{PT}$, thus DAF is expected to develop. 001AFM Mn$_2$Bi$_2$Te$_5$ is insulating with a full energy gap about $0.1$~eV, shown in Fig.~\ref{fig3}(a). Fig.~\ref{fig3}(b) shows the density of states (DOS) without spin-orbit coupling (SOC). The red states are the projected DOS of Mn $d$ orbitals. We can see that the occupied $3d^5$ states and unoccupied $3d^0$ states are well separated in energy, indicating a high spin state. The fat band structure calculated through projecting on Te and Bi $p_z$ orbitals is shown in Fig.~\ref{fig3}(c). A band inversion is clearly seen at the Fermi level around $\Gamma$ point, which indicates possible nontrivial topological property of Mn$_2$Bi$_2$Te$_5$ with nonzero spin Chern number~\cite{wangh2020}. However, different from $\mathcal{T}$-invariant TIs, by gradually increasing SOC strength $\lambda$ from zero, the energy gap of Mn$_2$Bi$_2$Te$_5$ first decreases to a minimum without closing because of the $m_5$ mass induced by AFM ordering, and then increases. The band inversion happens at energy gap minimum, which is around $0.9\lambda$ for Mn$_2$Bi$_2$Te$_5$ shown in Fig.~\ref{fig3}(d). We further confirm that the topological transition is robust to onsite $U$, see Fig.~\ref{fig3}(e).

\begin{figure}[b]
\begin{center}
\includegraphics[width=3.4in,clip=true]{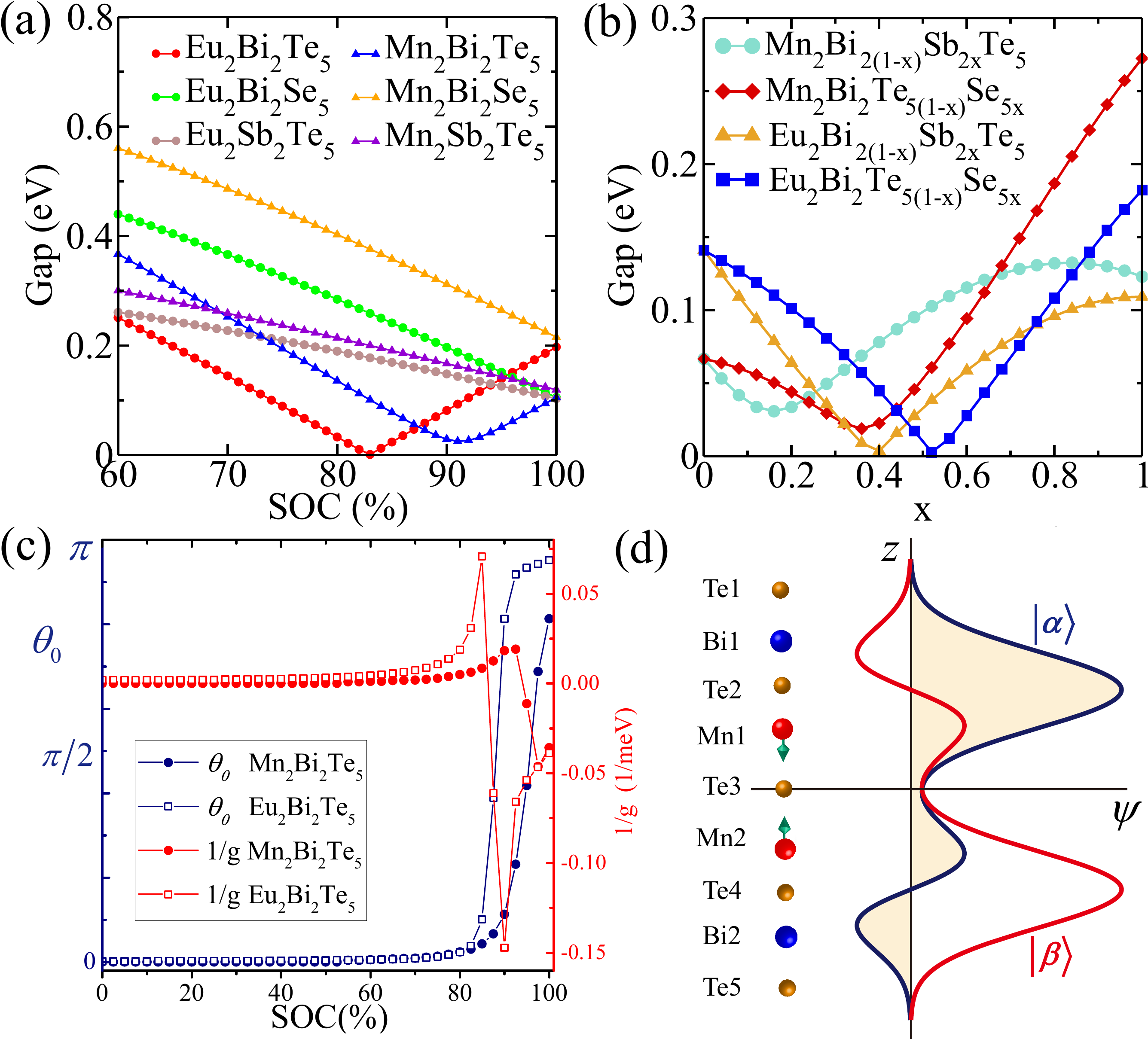}
\end{center}
\caption{(a) The evolution of energy gap at $\Gamma$ with $\lambda$. (b) The evolution of energy gap at $\Gamma$ with the doping strength. $x$ denotes the element substitution of Bi/Sb and Te/Se. For $x<x_c$ (minimum gap), the system has band inversion. (c) The static $\theta_0$ and $g$ of X$_2$Bi$_2$Te$_5$ vs $\lambda$, which are calculated from the effective $\mathbf{k}\cdot\mathbf{p}$ models. (d) Sketch of the wavefunction of states $|\alpha\rangle$ and $|\beta\rangle$ in each unit cell.}
\label{fig4}
\end{figure}

Fig.~\ref{fig4}(a) shows the evolution of energy gap as a function of $\lambda$ for 001AFM X$_2$A$_2$B$_5$ materials. Eu$_2$Bi$_2$Te$_5$ is topologically nontrivial with the minimum gap occuring at $\sim0.82\lambda$. While the gaps of X$_2$Sb$_2$Te$_5$, X$_2$Bi$_2$Se$_5$ and X$_2$Sb$_2$Se$_5$ monotonously decrease and do not arrive at a minimum, which indicates that they are topologically trivial due to reduced SOC compared to X$_2$Bi$_2$Te$_5$. Interestingly, one can effectively tune $\lambda$ and energy gap by element substitution (Bi/Sb, Te/Se) as shown in Fig.~\ref{fig4}(b), which further optimizes materials for large DAF. Each line has a minimum at certain $x_c$ indicating a band inversion for a topological transition.

\emph{Low-energy effective model.} To characterzie the low-energy and topological properties of Mn$_2$Bi$_2$Te$_5$, an effective Hamiltonian can be constructed. We start with the four low-lying states at $\Gamma$, which are $|P1_z^+,\uparrow(\downarrow)\rangle$ and $|P2_z^-,\uparrow(\downarrow)\rangle$. Without the SOC effect, the bonding state $|P1_z^+\rangle$ of two Bi layers stays above of the anti-bonding state $|P2_z^-\rangle$ of two Te layers (Te1 and Te5 in NLs), where the superscripts ``$+$'', ``$-$'' stand for the parity. The SOC effect pushes down $|P1_z^+,\uparrow(\downarrow)\rangle$ state and pushes up $|P2_z^-,\uparrow(\downarrow)\rangle$ state, leading to the band inversion and parity exchange. The symmetries of the system is three-fold rotation symmetry $C_{3z}$ and $\mathcal{PT}$ (Te3 as the center). In the basis of $(|P1^+_z,\uparrow\rangle,|P2^-_z,\uparrow\rangle,|P1^+_z,\downarrow\rangle,|P2^-_z,\downarrow\rangle,)$, the representation of the symmetry operations is given by $C_{3z}=\exp[i(\pi/3)\sigma^z]$ and $\mathcal{PT}=i\tau^z\sigma^y\mathcal{K}$ ($\mathcal{P}=\tau^z$, $\mathcal{T}=i\sigma^y\mathcal{K}$), where $\mathcal{K}$ is the complex conjugation operator, $\sigma^{x,y,z}$ and $\tau^{x,y,z}$ denote the Pauli matrices in the spin and orbital space, respectively. By requiring these symmetries and keeping only the terms up to quadratic order in $\mathbf{k}$, we obtain the following generic form of the effective Hamiltonian 
\begin{eqnarray}
\mathcal{H}(\mathbf{k}) &=& \epsilon_0(\mathbf{k})+m_4(\mathbf{k})\tau^z+m_5(\mathbf{k})\tau^y+A_1k_z\tau^x\sigma^z
\nonumber
\\
&&+\mathcal{A}_2(k_z)(k_x\tau^x\sigma^y-k_y\tau^x\sigma^x), 
\end{eqnarray}
where $\epsilon_0(\mathbf k)=C_0+C_1k_z^2+C_2(k_x^2+k_y^2)$, $m_4(\mathbf k)=M_0+M_1k_z^2+M_2(k_x^2+k_y^2)$, $m_5(\mathbf{k})=B_0+B_1k_z^2+B_2(k_x^2+k_y^2)$, and $\mathcal{A}_2(k_z)=A_2+A_3k_z$. It is very similar to the Dirac model in Eq.~(\ref{Dirac}), where $m_4(\mathbf{k})$ is the $\mathcal{T}, \mathcal{P}$-conserving mass term responsible for band inversion. $m_5(\mathbf{k})$ is $\mathcal{T}, \mathcal{P}$-breaking mass term from the 001AFM order. To see the microscopic origin of $m_5(\mathbf{k})$, we change the basis to $(|\alpha,\uparrow\rangle,|\alpha,\downarrow\rangle,|\beta,\uparrow\rangle,|\beta,\downarrow\rangle)$ with $|\alpha,\uparrow(\downarrow)\rangle=(1/\sqrt{2})(|P1_z^+,\uparrow(\downarrow)\rangle+|P2_z^-,\uparrow(\downarrow)\rangle)$ and $|\beta,\uparrow(\downarrow)\rangle=(1/\sqrt{2})(|P1_z^+,\uparrow(\downarrow)\rangle-|P2_z^-,\uparrow(\downarrow)\rangle)$. We have $\mathcal{P}|\alpha,\uparrow(\downarrow)\rangle=|\beta,\uparrow(\downarrow)\rangle$. The sketch of $|\alpha\rangle$ and $|\beta\rangle$ wavefunctions in NLs is shown in Fig.~\ref{fig4}(d). Physically, $|\alpha,\uparrow(\downarrow)\rangle$ and $|\beta,\uparrow(\downarrow)\rangle$ stand for states with the wavefunction majorly localized around Bi1 and Bi2 in the unit cell. By transforming $m_5\tau^y$ to the new basis, we see that it represents a staggered Zeeman field that points in the $+z(-z)$ direction on $|\alpha\rangle$ and $|\beta\rangle$. Since $|\alpha\rangle$ and $|\beta\rangle$ majorly overlaps with Mn1 and Mn2, respectively. This staggered Zeeman field is exactly generated by the AFM order on Mn atoms, with electron spins point along opposite $z$ directions on Mn1 and Mn2. Thus the AFM amplitude excitations further induce fluctuations of $\theta(\mathbf{r},t)$. The parameters are determined from first-principles calculations~\cite{supple}. We notice that $M_0<0$ and $M_1, M_2>0$, which correctly characterizes the band inversion around $\Gamma$ point. In the nonmagnetic state, $m_5(\mathbf{k})=A_3=0$, the system is a TI with a single gapless Dirac surface state. The direct consequence of $m_5(\mathbf{k})$ term is to open a gap of $2B_0$ in the surface-state spectrum, which is independent of the surface orientation. The $\theta_0$ and $1/g$ for 001 AFM X$_2$Bi$_2$Te$_5$ are calculated in Fig.~\ref{fig4}(c), where $1/|g|$ is largest around the topological phase transition consistent with the previous analysis.

\emph{Experimental proposals.} The DAF $\theta(\mathbf{r},t)$, once realized in Mn$_2$Bi$_2$Te$_5$, would induce a nonlinear magnetoelectric effect and can be measured by the nonlinear optical spectroscopy. A static magnetic field $\mathbf{B}_0$ and an a.c. electric field $\mathbf{E}(t)=\mathbf{E}_{\mathrm{ac}}\sin(\omega t)$ would excite $\delta\theta(\mathbf{r},t)\propto\sin(\omega t)\mathbf{E}_{\mathrm{ac}}\cdot\mathbf{B}_0$, which further induces a topological magnetization $\mathbf{M}_t\propto(\theta_0+\delta\theta(\mathbf{r},t))\mathbf{E}\propto a_1\sin(\omega t)+a_2\cos(2\omega t)$~\cite{supple}. The double frequency $2\omega$ is from the DAF~\cite{li2010}. The N\'{e}el order may also induce a second harmonic generation. The magnetic field, wavelength and temperature dependence would in principle to distinguish the DAF from static AFM order.

The nonzero $\delta\theta(\mathbf{r},t)$ in DAF leads to the magnetoelectric effects, which can be seen from the response equation by taking a variation in $\mathcal{S}_{\theta}$. Therefore the current density in the spatial space is~\cite{qi2008}
\begin{equation}\label{current}
\mathbf{j}(\mathbf{r},t) = \frac{1}{2\pi}\frac{e^2}{h}\left[\nabla\theta(\mathbf{r},t)\times\mathbf{E}+\partial_t\theta(\mathbf{r},t)\mathbf{B}\right].
\end{equation}
$\partial_t\equiv\partial/\partial t$. Here $\theta(\mathbf{r},t)=\theta_0+\delta\theta(\mathbf{r},t)$. For axion insulators, $\theta_0=\pi$ and $\delta\theta(\mathbf{r},t)=0$. The above action can also describe the axion electrodynamics of Weyl semimetals~\cite{armitage2018,grushin2012,son2012,zyuzin2012,goswami2013,vazifeh2013}, but $\theta(\mathbf{r},t)$ has its own dynamics in DAF from magnetic fluctuations. The first term in Eq.~(\ref{current}) is the anomalous Hall effect (AHE) induced by the spatial gradient of $\theta$. The second term in Eq.~(\ref{current}) is the CME~\cite{fukushima2008}, where an electric current is generated by magnetic fields from temporal gradient of $\theta$. Such an electric current is a polarization current $\mathbf{j}=\partial_t\mathbf{P}$ in insulators, where $\mathbf{P}$ is the charge polarization. In a static uniform magnetic field we have $\mathbf{P}=(\theta(\mathbf{r},t)/2\pi)(e^2/h)\mathbf{B}$. This is isotropic TME where charge polarization is induced by a parallel magnetic field. This is different from CME in Weyl semimetals, which only happens in nonequilibrium situations~\cite{armitage2018,zyuzin2012,vazifeh2013,zhou2013,pesin2015,zhong2016}. Also, CME will not happen in axion insulators with only static $\theta$.

The temporal dependent $\theta(t)$ can also be excited by the AFM resonance~\cite{baltz2018}. For example, a static magnetic field is applied along easy $z$ axis $\mathbf{B}=B\hat{\mathbf{z}}$ and a microwave is irradiated. The biparticle AFM spins are precessing around $z$ axis with the same resonant frequency $\omega_{\pm}=g\mu_BB\pm\sqrt{(2\omega_J+\omega_A)\omega_A}$, where $\omega_J$ and $\omega_A$ are the exchange field and anisotropy field, respectively~\cite{sekine2016}. The AFM order parameter now becomes $\mathbf{m}_{\pm}(t)\equiv(1/2)(\langle \mathbf{s}_{i\alpha}\rangle-\langle \mathbf{s}_{i\beta}\rangle)\approx n_0\hat{\mathbf{z}}+\delta\mathbf{m}_{\pm}e^{i\omega_{\pm}t}$, where $\mathbf{s}_{i\alpha}$ and $\mathbf{s}_{i\beta}$ is the sublattice spin. In a mean field theory of an interacting Hamiltonian, $m_5=-(2/3)Um^z$, where $U$ is the on-site repulsion~\cite{li2010}. Therefore, $\delta m^z_{\pm}(t)$ induces $\delta\theta(t)$, and from Eq.~(\ref{current}) one expects the CME with a polarization current~\cite{supple}.

It is worth mentioning that the sister compound MnBi$_2$Te$_4$, albeit a slightly different material, is an axion insulator with static $\theta=\pi$~\cite{zhang2019}. Though AFM resonance could induce AFM spins precessing, as long as the surface-states remain magnetically gapped, $\theta$ is still static to be $\pi$ due to conserved $\mathcal{P}$. Therefore no CME is expected to exist in MnBi$_2$Te$_4$.

In summary, the intrinsic van der Waals magnetic materials Mn$_2$Bi$_2$Te$_5$ family may provide the first experimental platform for DAF. We expect superlattice-like new magnetic TI such as Mn$_2$Bi$_2$Te$_5$/MnBi$_2$Te$_4$ and Mn$_2$Bi$_2$Te$_5$/Bi$_2$Te$_3$ with tunable exchange interactions and topological properties may be fabricated. This will further enrich the magnetic TI family and provide a new material platform for exotic topological phenomena from large DAF. The DAF predicted here may help to search for the elementary dark axion particle in high energy physics~\cite{marsh2019}.

\begin{acknowledgments}
J.W. acknowledge Jiang Xiao, Yayu Wang, Ke He, Naoto Nagaosa and Akihiko Sekine for helpful discussions. H.Z. is supported by the Fundamental Research Funds for the Central Universities 020414380149, the Natural Science Foundation of China under Grant No.~11674165 and No.~11834006, the Fok Ying-Tong Education Foundation of China under Grant No.~161006. J.W. is supported by the National Key Research Program of China under Grant No.~2016YFA0300703 and No.~2019YFA0308404, the Natural Science Foundation of China through Grant No.~11774065, Shanghai Municipal Science and Technology Major Project under Grant No. 2019SHZDZX04, the Natural Science Foundation of Shanghai under Grant Nos.~19ZR1471400. J.Z., D.W. and M.S. contributed equally to this work.
\end{acknowledgments}

\end{document}